\documentclass[a4paper]{article}
\pdfoutput=1
\usepackage[utf8]{inputenc}
\usepackage[T1]{fontenc}
\usepackage{fixltx2e}
\usepackage{graphicx}
\usepackage{longtable}
\usepackage{float}
\usepackage{wrapfig}
\usepackage{rotating}
\usepackage[normalem]{ulem}
\usepackage{amsmath}
\usepackage{textcomp}
\usepackage{marvosym}
\usepackage{wasysym}
\usepackage{amssymb}
\usepackage{hyperref}
\usepackage{graphicx}
\usepackage{color}
\usepackage{listings}
\hypersetup{colorlinks=true, linkcolor=blue}
\usepackage{units}
\usepackage{comment}
\usepackage{rotfloat}
\usepackage{a4wide}
\usepackage{ulem}
\usepackage{lmodern}
\usepackage{tabularx}
\usepackage{ltablex}
\usepackage{booktabs}
\usepackage{makeidx}
\makeindex

\renewcommand{\d}{{\rm d}}
\renewcommand{\vec}[1]{{\bf #1}}
\newcommand{\R}{{\mathbb{R}}}

\author{Johannes Willkomm}
\date{\today}
\title{Automatic differentiation of ODE integration}
\hypersetup{
  pdfkeywords={},
  pdfsubject={},
  pdfcreator={Emacs 25.2.2 (Org mode 8.2.10)}}
\begin{document}

\maketitle
\tableofcontents

\section{Introduction}
\label{sec-1}

In some use cases of automatic differentiation, is is desired to
differentiate a code list which contains calls to an ODE integration
routine, as for example the \textbf{ode23} solver in MATLAB or GNU
Octave. This function is one of a family of similar builtins, which
all have the same interface, such as for example \textbf{ode15s} or \textbf{ode45}.

In ADiMat \cite{Bischof2002CST,Willkomm2014ANU} this will currently
throw an error as the derivative of the builtin function \textbf{ode23} is
not yet specified. On the other hand some users have use cases where
this situation occurs and hence it is desirable that a method is
developed for handling such cases.

This paper is organised as follows: in Section \ref{s:odediff} we describe
the mathematical methods for computing the derivatives of the solution
of an ODE w.r.t. some parameters. In Section \ref{s:lotka} we introduce a
simple ODE system that serves as an illustrating example, the
well-known Lotka-Volterra equations, and show the analytic derivatives
for it. In Section \ref{s:adimat} we explain how the derivatives can be
computed with ADiMat, in particular when the use code uses the ODE
integration function \textbf{ode23}. We show how to construct working
substitution functions to propagate the derivative in both forward and
reverse mode, and also how support was added to ADiMat to compute the
Hessian of such user code in forward-over-reverse mode. Finally in
Section \ref{s:conclusion} we provide a set of conclusions.

\section{Derivatives of ODE solutions}
\label{sec-2}
\label{s:odediff}

Consider a simple ODE system of the form
\[
Y^\prime = f(t, Y)
\]
which describes the behaviour of $M$ time-dependent quantities $Y(t)
\in \R^M$ given an initial state $Y(t_0)$ at initial time $t_0 = 0$.

The most simple way to determine $Y(t, P)$ for times $t > t_0$ is by
integrating forward from the initial state using the well-known
explicit Euler method using small discrete time steps $\delta t$:
\[
Y(t_k) = Y(t_{k-1}) + \delta t Y^\prime(t_{k-1})
\]
with $t_k = k \delta t$ for $k=1,\dots,N$. In practice such ODE systems 
are routinely solved with more powerful methods such as the
Runge-Kutta schemes \cite{} of various orders.

Now it is often the case that the derivative of the ODE depends on a
set of parameters $P \in \R^K$
\[
Y^\prime = f(t, Y, P),
\]
and thus its solution can also be seen as a function of $P$, that is $Y =
Y(t,P)$. Then let us further assume that the derivatives of $Y$ w.r.t. the
parameters $P$ are of interest, i.e. the quantity $\frac{\d Y}{\d P}
\in \R^{M \times P}$. The general method to obtain the derivatives of
an ODE solution w.r.t. some parameters or the initial condition is
described in \cite{lutzl-2017}, which we will restate in the following
for our problem setup.

Since
\[ \frac{\d}{\d t} \frac{\d Y}{\d P} = \frac{\d f(t, Y, P)}{\d P}, \]
we get
\[ \frac{\d}{\d t} \frac{\d Y}{\d P} = \frac{\d f(t, Y, P)}{\d Y} \frac{\d Y}{\d P} + \frac{\d f(t, Y, P)}{\d P} \]
and by setting $V = \frac{\d Y}{\d P}$ see that we obtain an augmented ODE system
\begin{align}
Y^\prime &= f(t, Y, P) \label{e:aug1} \\
V^\prime &= f_Y(t, Y, P) \cdot V + f_P(t, Y, P) \label{e:aug2}
\end{align}
which can be integrated in the same manner as the original
system. Assuming that the initial values $Y(t_0)$ do not depend on the
parameters, the initial values $V(t_0)$ are set to zero.

When the derivatives of the solution $Y(t)$ w.r.t. the initial values
$Y(t_0)$ are of interest we get with a very similar reasoning for $W =
\frac{\d Y}{\d Y(t_0)}$ the additional equation
\begin{align}
W^\prime &= f_Y(t, Y, P) \cdot W. \label{e:aug3}
\end{align}
In this case the initial values $W(t_0)$ are set to the identity
matrix.

\section{Example: Differentiating the Lotka-Volterra ODE system}
\label{sec-3}
\label{s:lotka}

As an example we consider the Lotka-Volterra ODE system in Subsection
\ref{s:lotka_1}, derive the analytical derivatives for it in Subsection
\ref{s:lotkaderivs} and compute the derivatives with several different
approaches, including automatice differentiation and numerial methods
in Subsection \ref{s:calclotkaderivs}.

\subsection{The Lotka-Volterra ODE system}
\label{sec-3-1}
\label{s:lotka_1}

We consider a system of two ODEs describing the
population numbers of two interdependent species, for example rabbits
$Y_1$ and foxes $Y_2$; the well-known Lotka-Volterra
equations given by
\begin{align}
Y_1^\prime &= (\epsilon_1 - \gamma_1 Y_2) \cdot Y_1  \label{e:f1} \\
Y_2^\prime &= -(\epsilon_2 - \gamma_2 Y_1) \cdot Y_2   \label{e:f2}
\end{align}
with $P = [\epsilon_1, \gamma_1, \epsilon_2, \gamma_2]$ a set of four
positive real parameters \cite{lotka-1925,volterra-1931,wiki-en-lotka-volterra}.

\begin{figure}[htb]
\centering
\includegraphics[width=.9\linewidth]{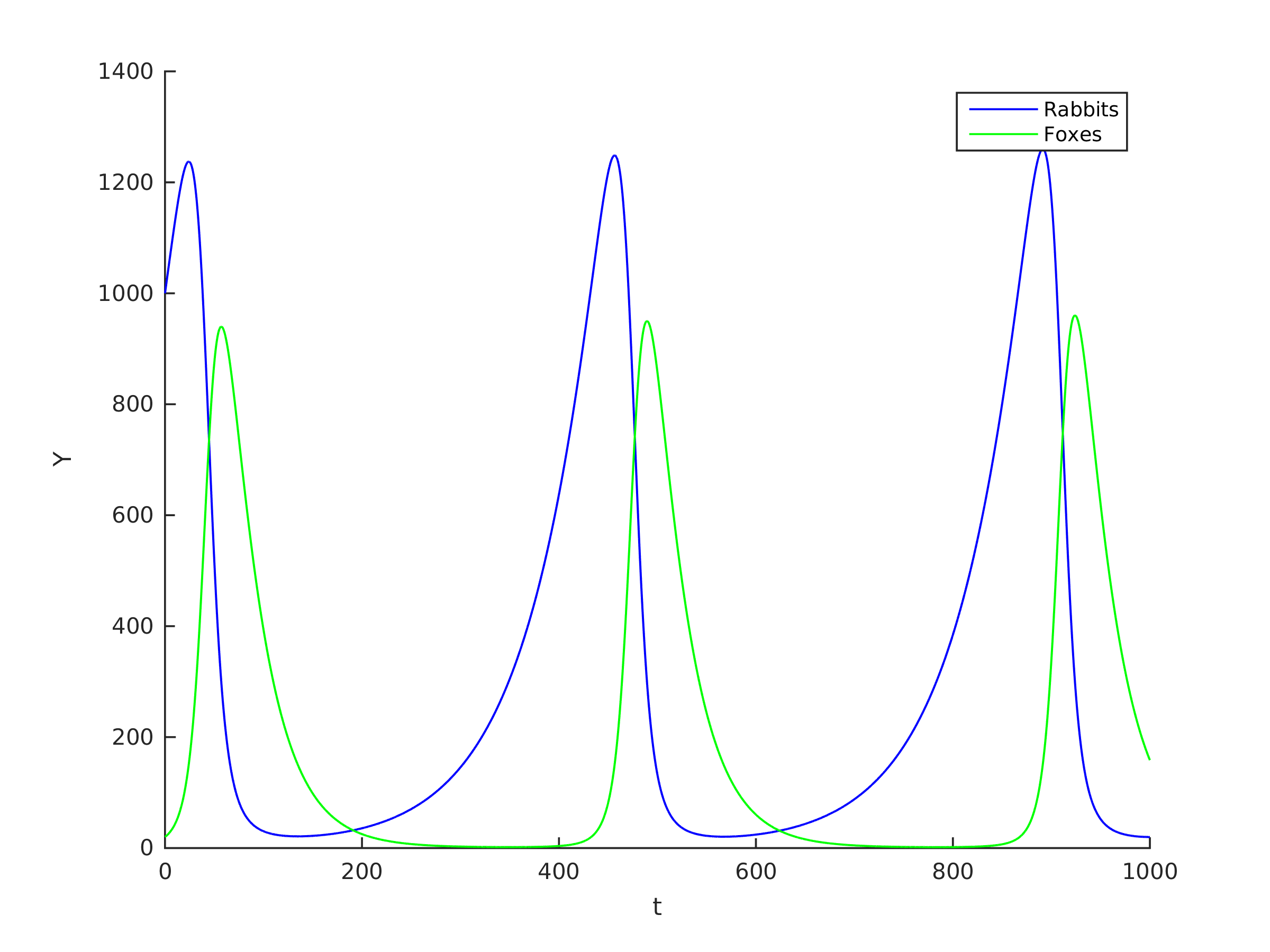}
\caption{\label{f:lvsol}An example solution of the Lotka-Volterra equations integrated with explicit Euler.}
\end{figure}

For our example we choose the parameters $P$ as given in the following table
\begin{center}
\begin{tabular}{rrrr}
$\epsilon$$_{\text{1}}$ & $\gamma$$_{\text{1}}$ & $\epsilon$$_{\text{2}}$ & $\gamma$$_{\text{2}}$\\
0.015 & 0.0001 & 0.03 & 0.0001\\
\end{tabular}
\end{center}
and set the initial conditions as follows:
\begin{center}
\begin{tabular}{rr}
Y$_{\text{1}}$(t$_{\text{0}}$) & Y$_{\text{2}}$(t$_{\text{0}}$)\\
1000 & 20\\
\end{tabular}
\end{center}
When we integrate the ODE in the time span $T = [t_0, t_{\rm end}] = [0, 10^3]$ with the
explicit Euler method with fixed time steps $\delta t = 0.1$ we obtain
the result shown in Figure \ref{f:lvsol}.

\subsection{The derivatives of the Lotka-Volterra ODE system}
\label{sec-3-2}
\label{s:lotkaderivs}

Now we apply the equations \eqref{e:aug1}--\eqref{e:aug3} to our example
problem. The derivative $f_Y$ is given by
\begin{align}
f_Y = \frac{\d f}{\d Y} &= \left(\begin{array}{cc} \epsilon_1 - \gamma_1 Y_2& -\gamma_1 Y_1 \\
                                               \gamma_2 Y_2 & -(\epsilon_2 - \gamma_2 Y_1) \end{array} \right),\label{e:dfy}
\end{align}
the derivative $f_P$ is
\begin{align}
f_P = \frac{\d f}{\d P} &= \left(\begin{array}{cccc}  Y_1 & -Y_1\cdot Y_2 &      0   &                    0  \\
                                                     0  &            0  &     -Y_2 &    Y_1 \cdot Y_2 \end{array} \right),\label{e:dfp}
\end{align}
we set the initial state $V(t_0)$ to zero
\begin{align*}
V(t_0) = \frac{\d Y(t_0)}{\d P} &= \left(\begin{array}{cccc}  0 & 0 & 0 & 0  \\
                                                              0 & 0 & 0 & 0 \end{array} \right),
\end{align*}
and the initial state $W(t_0)$ to the identity matrix
\begin{align*}
W(t_0) = \frac{\d Y(t_0)}{\d Y(t_0)} &= \left(\begin{array}{cc}  1 & 0  \\
                                                                 0 & 1 \end{array} \right).
\end{align*}

\begin{figure}[htb]
\centering
\includegraphics[width=.9\linewidth]{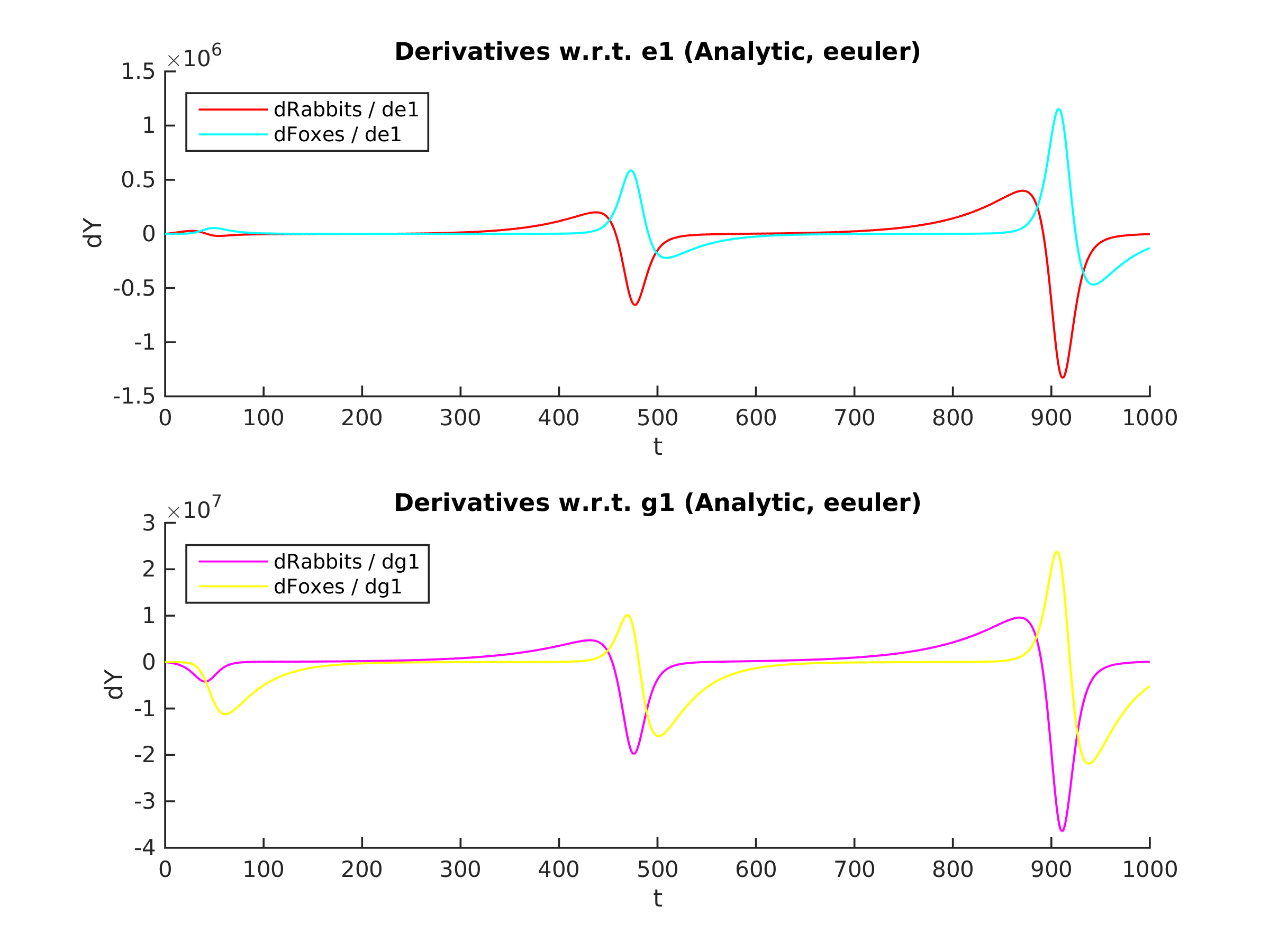}
\caption{\label{f:dlv_ana_euler1}The derivatives of the Lotka-Volterra equations w.r.t. the parameters $\epsilon_1$ and $\gamma_1$}
\end{figure}

\begin{figure}[htb]
\centering
\includegraphics[width=.9\linewidth]{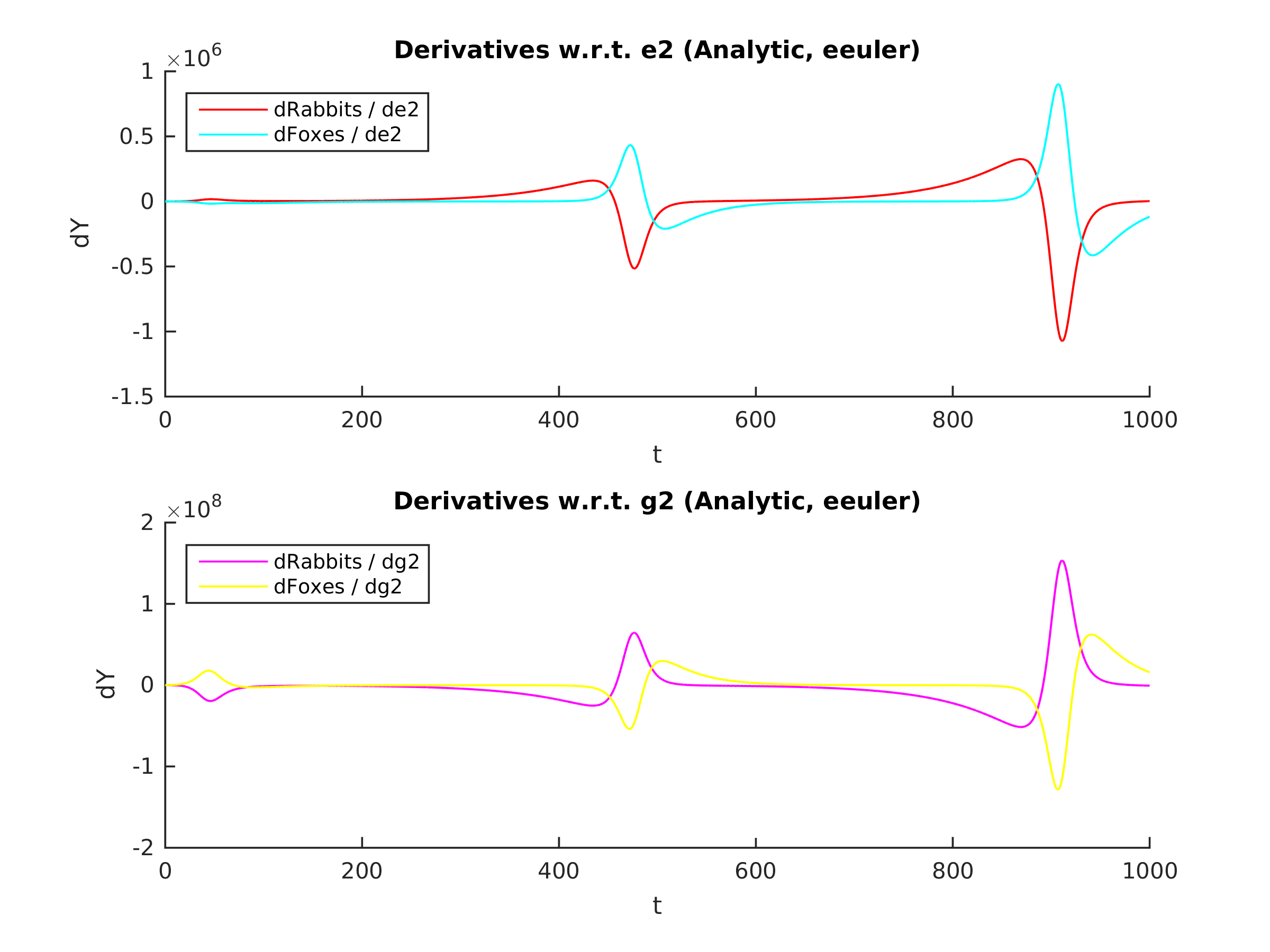}
\caption{\label{f:dlv_ana_euler2}The derivatives of the Lotka-Volterra equations w.r.t. the parameters $\epsilon_2$ and $\gamma_2$}
\end{figure}

\begin{figure}[htb]
\centering
\includegraphics[width=.9\linewidth]{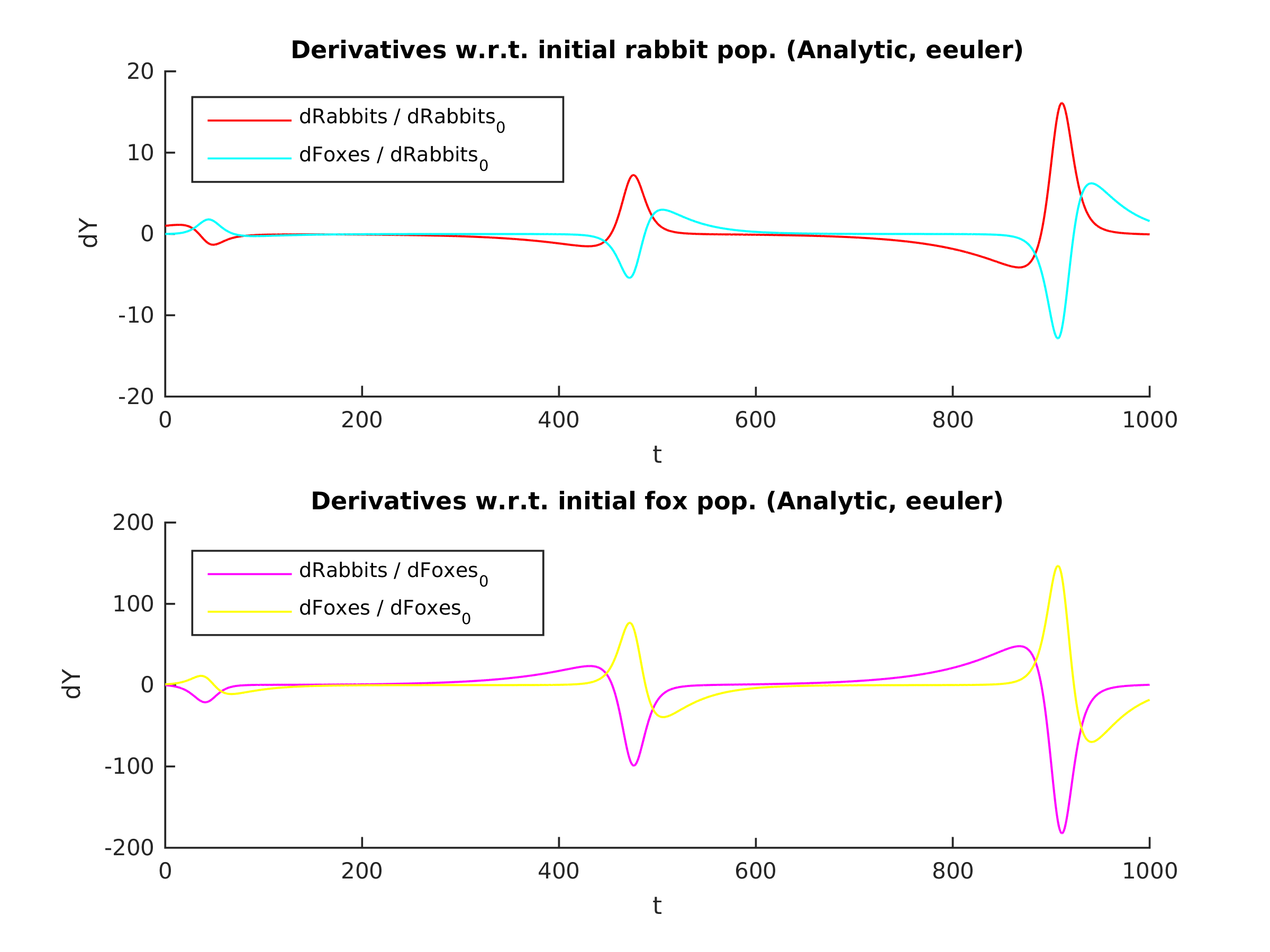}
\caption{\label{f:dlv_ana_euler3}The derivatives of the Lotka-Volterra equations w.r.t. the initial population numbers}
\end{figure}

Applying the explicit Euler method to the augmented system yields the
derivatives of the population numbers w.r.t. the four parameters over
time, as shown in Figures \ref{f:dlv_ana_euler1} and \ref{f:dlv_ana_euler2}, and
the derivatives w.r.t. the initial population numbers shown in Figure \ref{f:dlv_ana_euler3}.  Note that in
order to use the \textbf{ode23} solver the augmented ODE system has to be
cast into a single column vector, that is, the system is then solved
by integrating a composite state $\vec{x} = \left(\begin{array}{ccc} Y^T  & \left \lfloor V \right \rfloor^T   & \left \lfloor W \right \rfloor^T \end{array}\right)^T$ from the initial
state $\vec{x}(t_0)$ using the composite derivative
\begin{align}
 \vec{x}^\prime = \left(\begin{array}{c} Y^\prime
\\ \left \lfloor V^\prime  \right \rfloor 
\\ \left \lfloor W^\prime  \right \rfloor \end{array}\right) = \left(\begin{array}{c} f(t,Y,P)
\\ \left \lfloor \frac{\d f}{\d Y} V + \frac{\d f}{\d P}  \right \rfloor 
\\ \left \lfloor \frac{\d f}{\d Y} W  \right \rfloor \end{array}\right), \label{e:compsys}
\end{align}
where $\left \lfloor \cdot \right \rfloor$ shall denote the cast to a column vector.

We show the function \verb~dfode~ in Listing \ref{l:dfode}. This function defines
the composite ODE system as defined in \eqref{e:compsys}. Note how the
vector \verb~x~, which corresponds to $\vec{x}$, is split into the three parts
corresponding to $Y$, $V$, and $W$, and how the latter two are
reshaped to the correct shapes so that the matrix expressions $f_Y
\cdot V + f_P$ and $f_Y \cdot W$ can be evaluated correctly. The
functions \verb~fodep~, \verb~fodep_y~ and \verb~fodep_p~ are not shown, they
correspond exactly to $f$ from \eqref{e:f1}--\eqref{e:f2} and the
partial derivatives $f_Y$ from \eqref{e:dfy} and $f_P$ from
\eqref{e:dfp}, resp.

\lstset{language=matlab,label=l:dfode,caption={Function defining the augmented ODE system for the Lotka-Volterra equations},numbers=none}
\begin{lstlisting}
function dyp = dfode(t, x, p)
  y      =         x(1:2);
  dydp   = reshape(x(3:10),   [2,4]);
  dydy0  = reshape(x(11:end), [2,2]);
  dy     = fodep(t, y(1:2), p);
  ddydp  = fodep_y(t, y, p) * dydp + fodep_p(t, y, p);
  ddydy0 = fodep_y(t, y, p) * dydy0;
  dyp = [
      dy(:)
      ddydp(:)
      ddydy0(:)
  ];
\end{lstlisting}

Obviously we could also use automatic differentiation of the MATLAB
code implementing $f$ to obtain $f_Y$ and $f_P$. This approach yields
exactly the same results as the analytic derivatives. Here we could of
course use the ADiMat driver \verb~admDiffFor~, cf. \cite{Willkomm2014ANU},
but for such a small derivative being computed many times in a hot
loop the overhead of this convenience driver is just too large. This
driver can be made a little bit faster by setting the option
\uline{nochecks}. This option has the effect that neither will the source
transformation be done nor the transformed file be checked for
conformity. Thus, the user must first run the \verb~admDiffFor~ once
exactly as intended to let ADiMat produce the differentiated code
correctly. Then in the ODE function the option \uline{nochecks} is
added. However, the overall expense of the driver is still much too
high for this approach to be competitive here. Instead, the user just
take the time to code the appropriate manual invocations of the
differentiated code. In our test this results in a reduction of the
overall runtime by a factor of 10. We show the function \verb~adfode~ in
Listing \ref{l:adfode}, including a branch to select one of the two
variants.

\lstset{language=matlab,label=l:adfode,caption={The augmented ODE function using automatic differentiation with ADiMat},numbers=none}
\begin{lstlisting}
function dyp = adfode(t, x, p)
  y      =         x(1:2);
  dydp   = reshape(x(3:10),   [2,4]);
  dydy0  = reshape(x(11:end), [2,2]);
  useDriver = false;
  if useDriver
    [J, dy] = admDiffFor(@fodep, 1, t, y, p, admOptions('i', [2,3], 'nochecks', 1));
    dfdy = J(:,1:2);
    dfdp = J(:,3:end);
  else
    dfdy = zeros(2, 2);
    dfdp = zeros(2, 4);
    g_y = zeros(2, 1);
    g_p = zeros(4, 1);
    for k=1:2
      g_y(k) = 1;
      [g_dy, dy] = g_fodep(t, g_y, y, g_p, p);
      dfdy(:,k) = g_dy;
      g_y = zeros(2, 1);
    end
    for k=1:4
      g_p(k) = 1;
      [g_dy, dy] = g_fodep(t, g_y, y, g_p, p);
      dfdp(:,k) = g_dy;
      g_p = zeros(4, 1);
    end
  end
  ddydp  = dfdy * dydp + dfdp;
  ddydy0 = dfdy * dydy0;
  dyp = [
      dy(:)
      ddydp(:)
      ddydy0(:)
  ];
\end{lstlisting}

Another completely different approach is to numerically differentiate
the entire ODE integration, using either the finite difference
approximation (FD) method \cite{wiki-en-fd} or the complex step (CS)
method \cite{lyness1967numerical}. The latter method is well-known to
yield accurate derivatives of real analytic computations while the
typical accuracy of finite differences is at best half the machine
precision. Since both $f$ and the explicit Euler method are real
analytic we get the expected relative errors when comparing the four
different ways to evalute these derivatives, as shown in the
all-vs-all comparison shown in Table \ref{t:cross_eeuler}.

\begin{table}[htb]
\caption{\label{t:cross_eeuler}The relative error of the derivatives, computed with the four different methods, of the ODE integrated with the explicit Euler scheme, each compared against all the others}
\centering
\begin{tabular}{lrrr}
 & vs. AD & vs. FD & vs. CS\\
\hline
Analytic & 0 & 1.55324\,(-06) & 3.3805\,(-15)\\
AD &  & 1.55324\,(-06) & 3.3805\,(-15)\\
FD &  &  & 1.55324\,(-06)\\
\end{tabular}
\end{table}

Integration of the augmented ODE system with either analytic or AD
derivatives yields exactly the same result, while using the CS method
to differentiate through the integration of the original ODE system
yields very nearly the same result. The relative error in the divided
differences is a little more than the square root of the machine
precision, as expected.

\subsection{Integration of the original and augmented Lotka-Volterra ODE system with the Runge-Kutta method}
\label{sec-3-3}
\label{s:calclotkaderivs}

Things become a little more interesting when we use the more
sophisticated Runge-Kutta method with adaptive timesteps as it is
implemented in MATLAB or GNU Octave by the \textbf{ode23} builtin function.

The first thing to note is that when integrating the augmented system
the ODE integrator will generally choose different time step sizes
than for the original system. Hence it will yield result vectors of
different lengths, which makes it difficult to compare the different
differentiation methods. This can be circumvented by prescibing a
vector of time points to the integrator, instead of the tuple of start
and end points $T$, forcing it to return the solution at exactly the
given points. This is in particular also required when using finite
differences to approximate the derivative, because otherwise the
results are completely spurious. Appearently the perturbations in the
parameters are small enough to result in the same number of time
steps, but the output time points themselves do change.

\begin{figure}[htb]
\centering
\includegraphics[width=.9\linewidth]{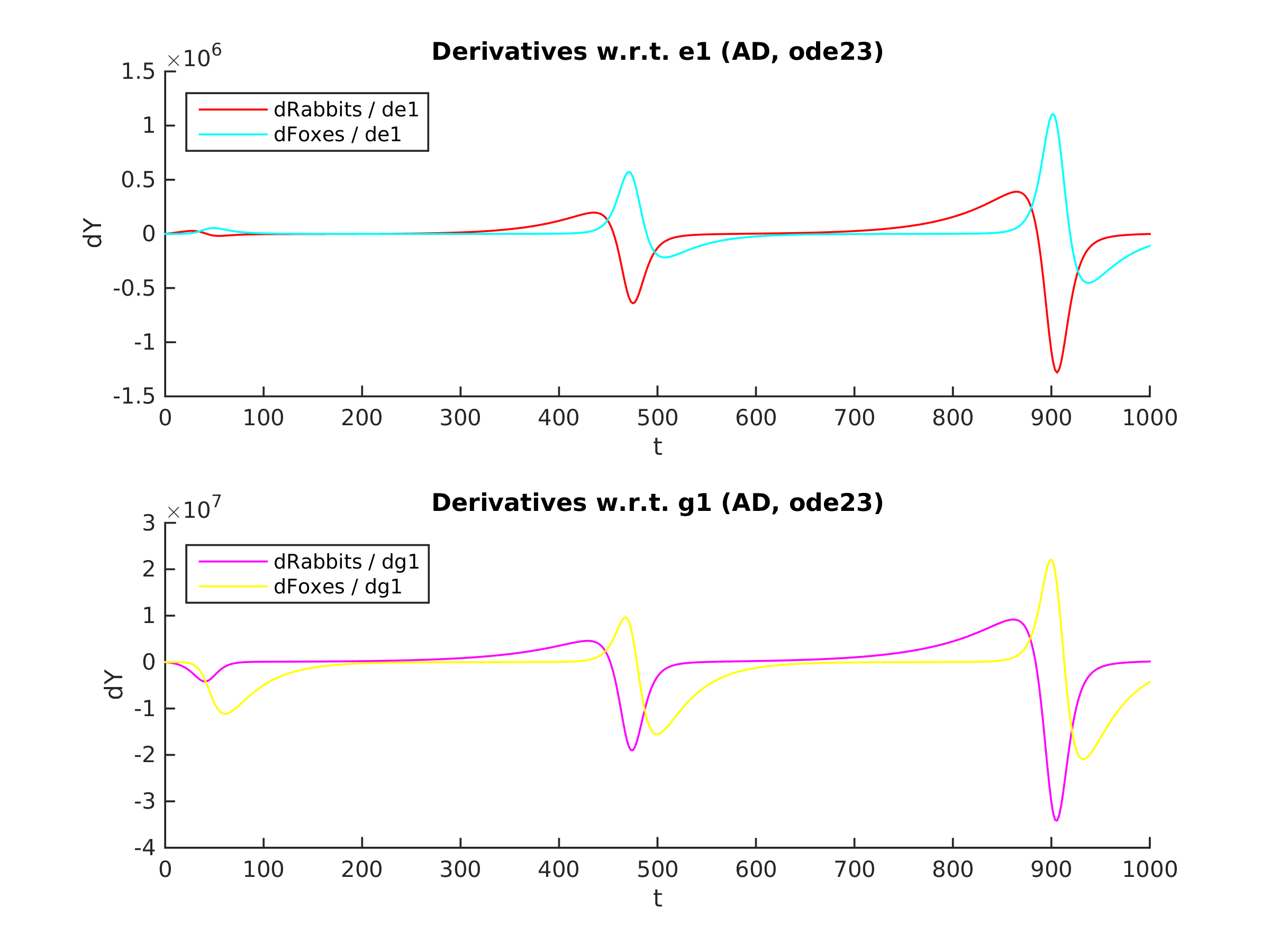}
\caption{\label{f:dlv_ad_ode23}The results of the complex step method applied to the integration of the Lotka-Volterra equations with the ode23 solver, w.r.t. the parameters $\epsilon_2$ and $\gamma_2$}
\end{figure}

\begin{figure}[htb]
\centering
\includegraphics[width=.9\linewidth]{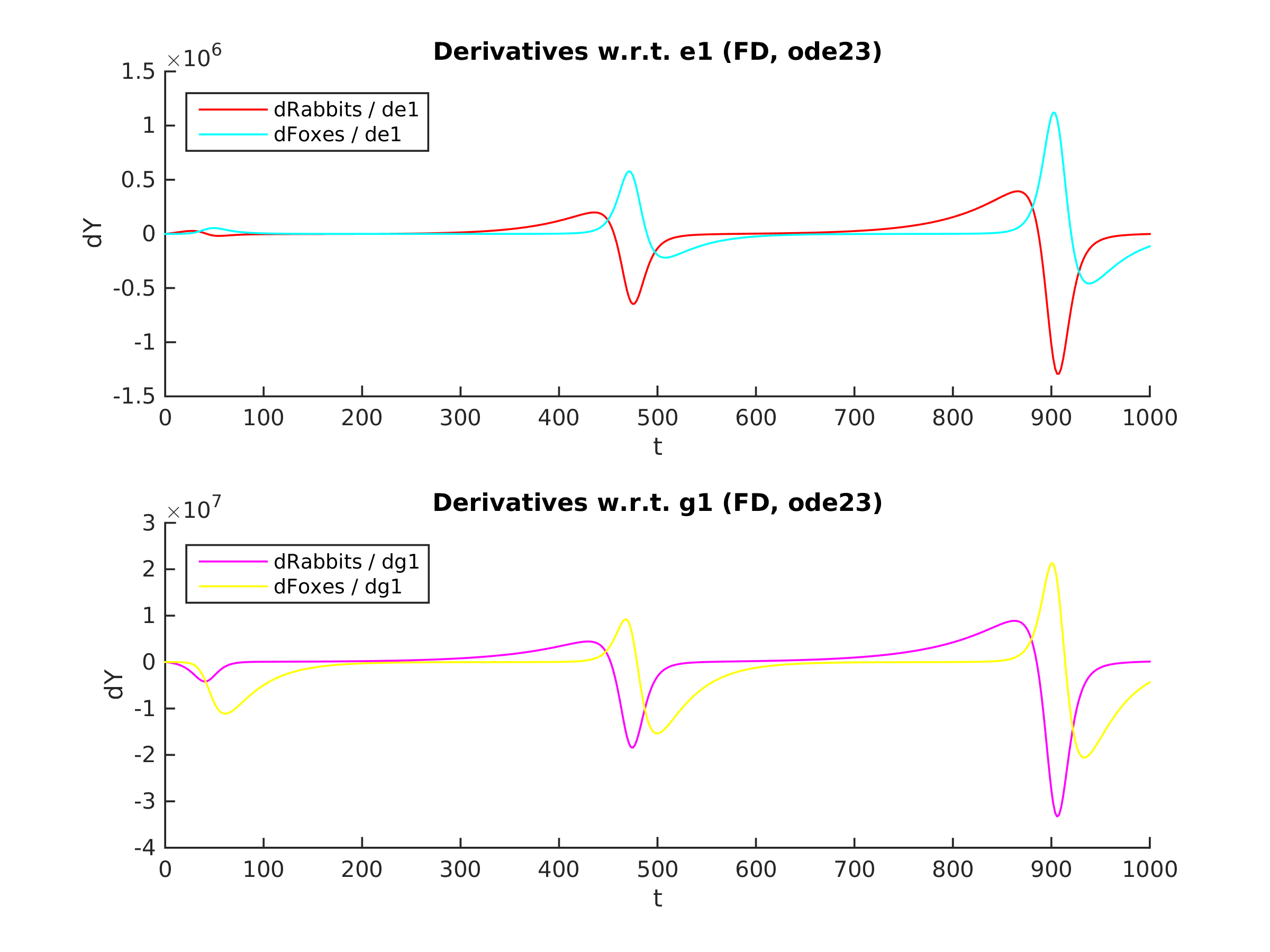}
\caption{\label{f:dlv_fd_ode23}The finite difference approximation to the derivative w.r.t. the parameters $\epsilon_1$ and $\gamma_1$ of integrating the Lotka-Volterra equations with the ode23 solver}
\end{figure}

\begin{figure}[htb]
\centering
\includegraphics[width=.9\linewidth]{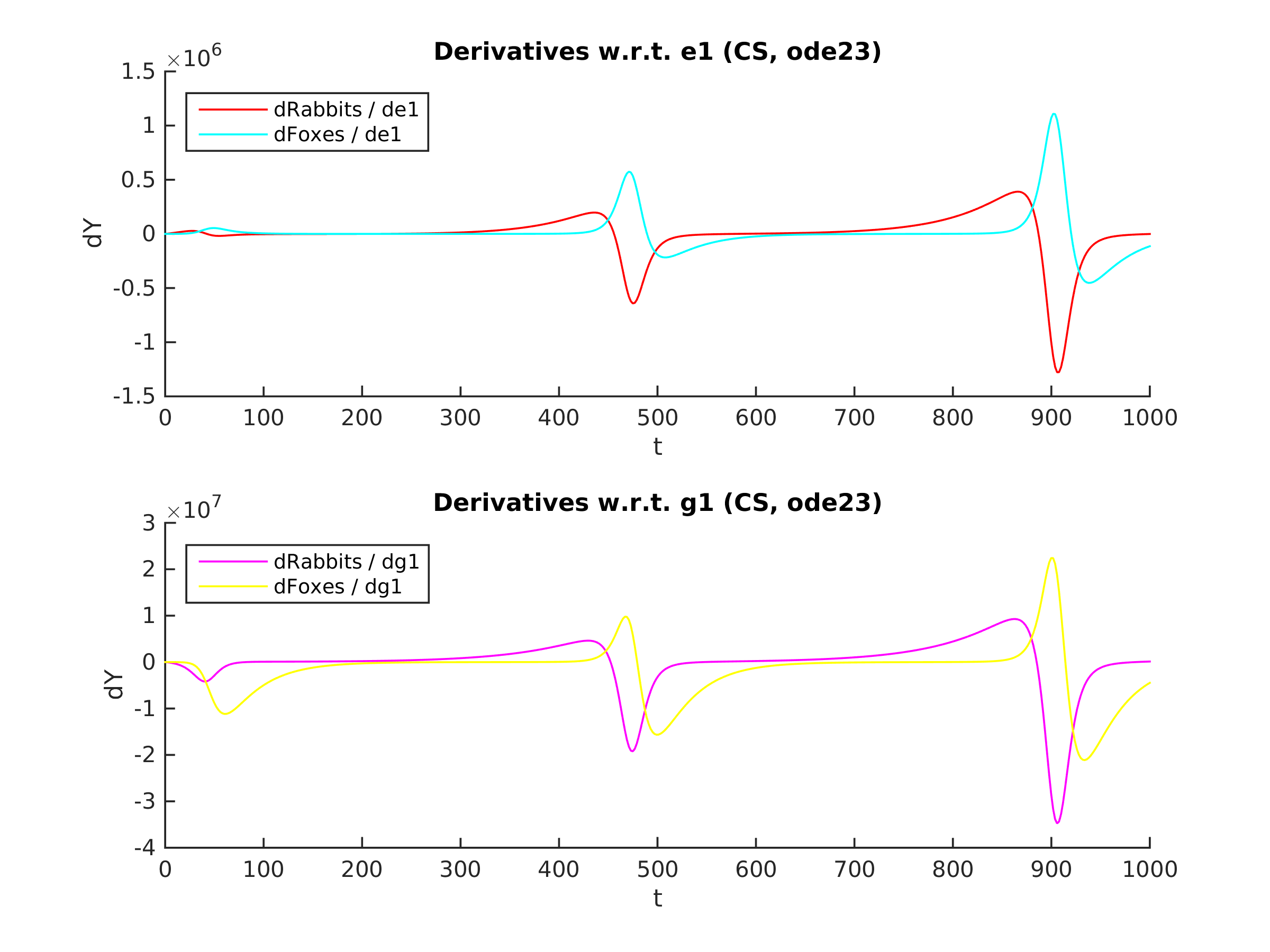}
\caption{\label{f:dlv_cs_ode23}The results of the complex step method applied to the integration of the Lotka-Volterra equations with the ode23 solver, w.r.t. the parameters $\epsilon_2$ and $\gamma_2$}
\end{figure}

The results of the augmented system with either analytic or AD
derivatives are visually very similar to those with the explicit Euler
method, shown in the previous subsection. We show the AD derivative
w.r.t. the first two parameters in Figure \ref{f:dlv_ad_ode23}, we show the
FD approximation w.r.t. the first two parameters in the Figure
\ref{f:dlv_fd_ode23}, and the CS method derivative w.r.t. the same
parameters in Figure \ref{f:dlv_cs_ode23}.

However, the relative error all-vs-all relative error comparison shown
in Table \ref{t:cross_ode23} reveals that the FD and CS methods are
appearently yielding results which are substantially different from
the analytic and AD derivatives in this case.

\begin{table}[htb]
\caption{\label{t:cross_ode23}The relative error of the derivatives w.r.t. $P$, computed with the four different methods, of the ODE integrated with the ode23 solver, each compared against all the others}
\centering
\begin{tabular}{lrrr}
 & vs. AD & vs. FD & vs. CS\\
\hline
Analytic & 0 & 0.0398952 & 0.0411807\\
AD &  & 0.0398952 & 0.0411807\\
FD &  &  & 0.0112255\\
\end{tabular}
\end{table}

The runtimes with all four methods, using explicit Euler or \textbf{ode23}
are given in Table \ref{t:runtimes}. While using AD with the manually
written invocations of the differentiated code is slightly slower than
using analytic derivatives, it is about twice as fast as using the FD
or CS method.

\begin{table}[htb]
\caption{\label{t:runtimes}Runtimes of the augmented ODE equations with the four different methods to compute the derivatives, in seconds}
\centering
\begin{tabular}{lrrrr}
 & Analytic & AD & FD & CS\\
\hline
Expl. Euler & 0.845753 & 1.86251 & 3.04297 & 2.76348\\
\textbf{ode23} & 0.208587 & 0.318961 & 0.635216 & 0.600115\\
\end{tabular}
\end{table}

\section{Computing derivatives of ODE integrations with ADiMat}
\label{sec-4}
\label{s:adimat}

Although ADiMat is a relatively mature AD tool which supports a
substantial set of MATLAB builtin functions, due to the sheer number
of these it often occurs that a particular user code makes use of a
builtin that is not yet supported. An example for this is the \textbf{ode23}
ODE solver which we use here to solve the Lotka-Volterra equations. In
this section we want to show how to insert a manually derived
derivative into a larger piece of differentiated code.

Consider as an example the function \verb~fmain~ shown in Listing
\ref{l:fmain}. In order to facilitate this task it is most convinient to
place the call to \textbf{ode23} into a separate function, so as to isolate
it as much as possible. For this purpose we define the function
\verb~calcode~, which is shown in Listing \ref{l:calcode}.

\lstset{language=matlab,label=l:fmain,caption={Main function to be differentiated with ADiMat, which makes indirect use of the unsupported builtin \textbf{ode23}, via the sub function calcode},numbers=none}
\begin{lstlisting}
function z = fmain(y0, p, ts)
%ADiMat BMFUNC [$$1,$$2]=calcode($1,$2,$3) DIFFTO [$$@1,$$1,$$@2,$$2]=g_calcode($@1,$1,$@2,$2,$3)
  p2 = p ./ 2;
  [t1, y1] = calcode(y0, p, ts);
  [t2, y2] = calcode(y0, p2, ts);
  z = sum(y1(end,:) + y2(end,:));
\end{lstlisting}

\lstset{language=matlab,label=l:calcode,caption={Function calcode is used to isolate the call to the unsupported builtin ode23},numbers=none}
\begin{lstlisting}
function [t, yt] = calcode(y0, p, ts)
  [t, yt] = eeuler(@(t, y) fodep(t, y, p), ts, y0);
\end{lstlisting}

Then we differentiate the function \verb~fmain~ as normally in forward
mode. ADiMat will differentiate \verb~calcode~ as well and complain about
the unsupported builtin \textbf{ode23}. As a result the generated function
\verb~g_calcode~ will be invalid. However, by looking at the signature of
\verb~g_calcode~ we know how a manually created replacement function should
look like, as shown in Listing \ref{l:g_calcode_1}.

\lstset{language=matlab,label=l:g_calcode_1,caption={Signature of the invalid differentiated function for calcode generated by ADiMat},numbers=none}
\begin{lstlisting}
function [t, g_y, y] = g_calcode(y0, g_p, p, ts)
\end{lstlisting}

For the reverse mode, the approach is a little bit trickier. We delete
the file \verb~calcode.m~ before launching the differentiation. This will
lead ADiMat to creating a generic call to an adjoint function. Of
course we restore \verb~calcode.m~ afterwards. For this to work we have to
declare the identifier \verb~calcode~ to ADiMat using the \verb~BMFUNC~
directive \cite{bbv:mac}.

Next we prepare two functions which compute the correct derivatives
when called from the differentiated code of \verb~fmain~. This requires a
little bit of insight into how ADiMat works, so a few comments are in
order. The general approach is to obtain in any conceivable way the
local Jacobian $J$ or partial derivative of the function \verb~calcode~. To
this end we can set up the augmented ODE system
\eqref{e:aug1}--\eqref{e:aug2} using either analytic or AD
derivatives. In the latter case we would effectively use ADiMat in a
hierarchical, two-tier fashion.

Then, in forward mode, we reshape in incoming derivative to a column
vector and multiply it from the left with $J$. Afterwards we reshape
it to the same shape as the function result of \verb~calcode~. This is code
is placed in a file \verb~fm_calcode.m~ which is shown in Listing
\ref{l:fm_calcode}. Note one particularity: since \verb~g_calcode~ also returns
the function results of \verb~calcode~ we can replace these by extracting
the corresponding bits of the augmented ODE system. Remember however
that the result vectors have different lengths when solving the
augmented ODE. In this case, due what is actually done with these
results in \verb~fmain~ or \verb~g_fmain~ we get way with this substitution. In
other cases this may not be the case. We will see how to handle that
when we consider the reverse mode.

\lstset{language=matlab,label=l:fm_calcode,caption={FM substitution function for calcode, using the augmented ODE system with analytic derivatives to compute the derivative},numbers=none}
\begin{lstlisting}
function [t, g_y, y] = g_calcode(g_y0, y0, g_p, p, ts)
  dydp_0 = zeros(2, 4);
  dydy0_0 = eye(2);
  yp0 = [y0(:)
	 dydp_0(:)
	 dydy0_0(:)];
  [t, ypt] = ode23(@(t, yp) dfode(t, yp, p), ts, yp0);
  no = length(t);
  y   = ypt(:,1:2);
  Jp  = reshape(ypt(:,3:10), [2.*no, 4]);
  Jy0 = reshape(ypt(:,11:end), [2.*no, 2]);
  g_y = Jy0 * g_y0(:) + Jp * g_p(:);
  g_y = reshape(g_y, size(y));
\end{lstlisting}

In reverse mode, we reshape in incoming adjoint to a row vector and
multiply it from the right with $J$. Then we reshape the result to the
same shape as the function parameter of which we require the adjoint,
in this case \verb~p~. This is code is placed in a file \verb~rm_calcode.m~
which shown in Listing \ref{l:rm_calcode}. Here we run into the problem of
the changing number of integrations steps that we mentioned. The
function \verb~rm_calcode.m~ receives the adjoint of \verb~y~, and so we are
stuck with its size and need to create a Jacobian of conforming
size. To this end, we have to run the original ODE system first,
simply to get the vector of time steps. Of course this is only
necessary when the time argument \verb~ts~ to the \verb~calcode~ function is
really a time span, i.e. a vector of length two, and not a vector of
time points already. Obviously the second variant is preferable.

\lstset{language=matlab,label=l:rm_calcode,caption={RM substitution function for calcode, using the augmented ODE system with analytic derivatives to compute the derivative},numbers=none}
\begin{lstlisting}
function [a_y0 a_p] = a_calcode_110(y0, p, ts, a_y)
  if length(ts) == 2
    % timespan given: run ODE integration to get resulting time points
    [t, yt] = calcode(y0, p, ts);
  else
    t = ts;
  end
  dydp_0 = zeros(2, 4);
  dydy0_0 = eye(2);
  yp0 = [y0(:)
	 dydp_0(:)
	 dydy0_0(:)];
  % run augmented ODE integration with same time points as
  % non-augmented integration
  [t_alt, ypt] = ode23(@(t, yp) dfode(t, yp, p), t, yp0);
  no = length(t);
  assert(no == length(t_alt));
  y   =         ypt(:,1:2);
  Jp  = reshape(ypt(:,3:10), [2.*no, 4]);
  Jy0 = reshape(ypt(:,11:end), [2.*no, 2]);
  a_p  = a_y(:).' * Jp;
  a_y0 = a_y(:).' * Jy0;
  a_p  = reshape(a_p, size(p));
  a_y0 = reshape(a_y0, size(y0));
\end{lstlisting}

The entire process to produce and then run a valid function \verb~g_fmain~
which calls the manually prepared code in Listing \ref{l:fm_calcode} can be
expressed with the following MATLAB commands:
\lstset{language=matlab,label= ,caption= ,numbers=none}
\begin{lstlisting}
r_f = admTransform(@fmain,admOptions('i', [1,2], 'mode', 'F'))
copyfile('fm_calcode.m', 'g_calcode.m');
J_f = admDiffFor(@fmain, 1, y0, p, ts, admOptions('i', [1,2],'nochecks',1))
\end{lstlisting}
For the reverse mode, the following MATLAB commands will produce and
then run a valid function \verb~a_fmain~ which calls the manually prepared
code in Listing \ref{l:rm_calcode}:
\lstset{language=matlab,label= ,caption= ,numbers=none}
\begin{lstlisting}
if exist('calcode')
  movefile('calcode.m', '_calcode.m');
  clear calcode
end
r_r = admTransform(@fmain,admOptions('i', [1,2], 'mode', 'r'))
movefile('_calcode.m', 'calcode.m');
copyfile('rm_calcode_110.m', 'a_calcode_110.m');
J_r = admDiffRev(@fmain, 1, y0, p, ts, admOptions('i', [1,2],'nochecks',1))
\end{lstlisting}

The runtimes of derivative evaluations with AD in FM and RM are
compared to those with the FD and CS methods in Table \ref{t:runtimes_main}.

\begin{table}[htb]
\caption{\label{t:runtimes_main}Runtimes of the example function differentiated with ADiMat in FM, RM and with the FD and CS methods, in seconds}
\centering
\begin{tabular}{lrrrr}
 & FM & RM & FD & CS\\
\hline
Expl. Euler & 1.85121 & 2.39589 & 5.76037 & 5.35389\\
\textbf{ode23} & 2.06182 & 0.571279 & 1.11894 & 0.734798\\
\end{tabular}
\end{table}

\subsection{Evaluating the Hessian in forward-over-reverse mode}
\label{sec-4-1}

ADiMat can compute the Hessian matrix of a function using
forward-over-reverse mode, through its driver function
\textbf{admHessian}. In this case the code differentiated in reverse mode,
which is exactly the same function \verb~a_fmain~ that we created in the
previous section, is being run with arguments that are of a special
class \verb~tseries2~ with overloaded operators (OO) which propagate
derivatives -- more precisely, truncated Taylor coefficients -- in
forward mode. Attempting to use this to differentiate \verb~fmain~ will
fail however, with an error being thrown by \textbf{ode23}. It appears it is
not allowed to run the \textbf{ode23} solver with any other type than single
or double floats in MATLAB. This is the case even when the OO type
enters the picture only via the additional parameter \verb~p~ used by the
function handle given to the solver as the first argument, that is,
when we differentiate w.r.t. to $P$ only, but not w.r.t. $Y(t_0)$.

Hence it is required to add a method \textbf{ode23} to the OO class and
ensure that the call is dispatched to it. This would will only happen
when one of the immediate arguments to the call is of the OO
type. That means, one must compute the derivative w.r.t. \verb~y0~. Then it
is possible, albeit by means of some quite advanced MATLAB hacking, to
analyse the function handle passed, extract the OO value \verb~p~ from its
closure, or workspace, and construct a new one which, firstly, does
not have an OO type in the closure and, secondly, represents the
augmented ODE system which can be used to calculate the derivatives.

This derivative can then be used to propagate the derivatives to the
output OO value, just as in the FM example shown before. This approach
will only yield the first order derivatives, which is a limitation for
the class \verb~tseries2~, which in general can propagate Taylor coeffients
truncated at a finite but arbitrary order. However, computing the
first derivative in \verb~tseries2~ is enough to evaluate the Hessian since
the second order derivative is obtained by virtue of the reverse mode
code.

The runtime of the Hessian evaluation with \verb~admHessian~ is compared to
those with the second order FD method in Table \ref{t:runtimes_hess}.

\begin{table}[htb]
\caption{\label{t:runtimes_hess}Runtimes of the evaluation of the Hessian of example function with ADiMat and with the method, in seconds}
\centering
\begin{tabular}{lrr}
 & FM-over-RM & FD\\
\hline
Expl. Euler & 21.398 & 3.89455\\
\textbf{ode23} & 2.80302 & 7.81168\\
\end{tabular}
\end{table}

\section{Conclusion}
\label{sec-5}
\label{s:conclusion}

In this report we showed several methods to calculate the derivatives
of an ODE integration. In particular we showed how to use ADiMat to
differentiate a larger code that calls the MATLAB ODE solver \textbf{ode23}
in some sub function as part of its overall computations. This works
with both the forward and reverse mode of ADiMat but currently
requires some manual intervention. Obviously a continuation of these
results would look into discerning a method that automates the entire
process. To get this done correctly, however, will propably require
that ADiMat first supports lambda expressions, which is most likely
still a substantial development effort.

Another result that we obtain is that both the FD and CS methods can
be used to calculate the derivatives of a simple explicit Euler
scheme, however both fail when attempting to differentiate the \textbf{ode23}
solver. The exact reason should be investigated further, since it is
not usual that AD and FD derivatives deviate by more than the
to-be-expected inaccuracy. However, it is possible to give examples
where FD approximation can fare almost arbitrarily bad. Also, the fact
that we appearently get the correct derivatives when using the simple
explicit Euler scheme seems to show that we are indeed doing the
correct thing when setting up the augmented ODE system.

The Hessian matrix can also be computed by means of the
forward-over-reverse mode using the operated overloading Taylor
propagation class in ADiMat. The implementation of the \verb~ode23~ method
added to the class for this purpose shows the way to go for a full
support for the differentiation of ODE integration builtins with
ADiMat.

\bibliographystyle{acm}
\bibliography{bib}

\begin{thebibliography}{1}

\bibitem{Bischof2002CST}
{\sc Bischof, C.~H., B{\"u}cker, H.~M., Lang, B., Rasch, A., and Vehreschild,
  A.}
\newblock Combining source transformation and operator overloading techniques
  to compute derivatives for {MATLAB} programs.
\newblock In {\em Proceedings of the Second {IEEE} International Workshop on
  Source Code Analysis and Manipulation ({SCAM} 2002)\/} (Los Alamitos, CA,
  USA, 2002), IEEE Computer Society, pp.~65--72.

\bibitem{bbv:mac}
{\sc Bischof, C.~H., B{\"u}cker, H.~M., and Vehreschild, A.}
\newblock {A Macro Language for Derivative Definition}.
\newblock In {\em Automatic Differentiation: {A}pplications, Theory, and
  Implementations}, H.~M. B{\"u}cker, G.~F. Corliss, P.~D. Hovland, U.~Naumann,
  and B.~Norris, Eds., vol.~50 of {\em Lecture Notes in Computational Science
  and Engineering}. Springer, Berlin, 2005, pp.~181--188.

\bibitem{lotka-1925}
{\sc Lotka, A.~J.}
\newblock {\em Elements of Physical Biology}.
\newblock Williams \& Wilkins Company, Baltimore, 1925.
\newblock Accessed: 2018-01-29.

\bibitem{lutzl-2017}
{\sc LutzL}.
\newblock {Derivative of an ODE marching algorithm such as Euler or Runge
  Kutta}.
\newblock
  \url{https://math.stackexchange.com/questions/2119506/derivative-of-an-ode-marching-algorithm-such-as-euler-or-runge-kutta},
  2017.
\newblock Accessed: 2018-01-28.

\bibitem{lyness1967numerical}
{\sc Lyness, J., and Moler, C.}
\newblock Numerical differentiation of analytic functions.
\newblock {\em SIAM Journal on Numerical Analysis 4}, 2 (1967), 202--210.

\bibitem{volterra-1931}
{\sc Volterra, V.}
\newblock {\em {Leçons sur la Théorie Mathématique de la Lutte pour la
  Vie}}.
\newblock Gauthier-Villars, 1931.

\bibitem{wiki-en-fd}
{\sc {Wikipedia contributors}}.
\newblock Finite difference.
\newblock \url{https://en.wikipedia.org/wiki/Finite_difference}, 2018.
\newblock Accessed: 2018-01-31.

\bibitem{wiki-en-lotka-volterra}
{\sc {Wikipedia contributors}}.
\newblock Lotka–volterra equations.
\newblock \url{https://en.wikipedia.org/wiki/Lotka-Volterra_equations}, 2018.
\newblock Accessed: 2018-01-29.

\bibitem{Willkomm2014ANU}
{\sc Willkomm, J., Bischof, C.~H., and B{\"u}cker, H.~M.}
\newblock A new user interface for {ADiMat}: {T}oward accurate and efficient
  derivatives of {M}atlab programs with ease of use.
\newblock {\em International Journal of Computational Science and Engineering
  9}, 5/6 (2014), 408--415.

\end{thebibliography}
\end{document}